\PassOptionsToPackage{unicode}{hyperref}
\PassOptionsToPackage{hyphens}{url}
\documentclass[
]{article}
\usepackage{amsmath,amssymb}
\usepackage{lmodern}
\usepackage{iftex}
\ifPDFTeX
  \usepackage[T1]{fontenc}
  \usepackage[utf8]{inputenc}
  \usepackage{textcomp} 
\else 
  \usepackage{unicode-math}
  \defaultfontfeatures{Scale=MatchLowercase}
  \defaultfontfeatures[\rmfamily]{Ligatures=TeX,Scale=1}
\fi
\IfFileExists{upquote.sty}{\usepackage{upquote}}{}
\IfFileExists{microtype.sty}{
  \usepackage[]{microtype}
  \UseMicrotypeSet[protrusion]{basicmath} 
}{}
\makeatletter
\@ifundefined{KOMAClassName}{
  \IfFileExists{parskip.sty}{%
    \usepackage{parskip}
  }{
    \setlength{\parindent}{0pt}
    \setlength{\parskip}{6pt plus 2pt minus 1pt}}
}{
  \KOMAoptions{parskip=half}}
\makeatother
\usepackage{xcolor}
\ifLuaTeX
  \usepackage{selnolig}  
\fi
\IfFileExists{bookmark.sty}{\usepackage{bookmark}}{\usepackage{hyperref}}
\IfFileExists{xurl.sty}{\usepackage{xurl}}{} 
\hypersetup{
  pdftitle={The big challenge for livestock genomics is to make sequence data pay},
  hidelinks,
  pdfcreator={LaTeX via pandoc}}

\usepackage{authblk}

\title{The big challenge for livestock genomics is to make sequence data
    pay}
\author[1]{Martin Johnsson}
\affil[1]{Department of Animal Breeding and Genetics,
    Swedish University of Agricultural Sciences, Box 7023, 750 07 Uppsala, Sweden}

\date{2023-07-24}

\begin{document}
\maketitle

\hypertarget{introduction}{%
\section{Introduction}\label{introduction}}

This paper will argue that one of the biggest challenges for livestock
genomics is to make whole-genome sequencing and functional genomics
applicable to breeding practice. It discusses potential explanations for
why it is so difficult to consistently improve the accuracy of genomic
prediction by means of whole-genome sequence data, and three potential
attacks on the problem\footnote{I was invited to a special issue about
  potential future developments in the field, major accomplishments and
  what needs to be done to move forward. This is my response. While I
  greatly appreciate and respect the people who invited me, I am not too
  fond of the publisher and did not have the time to meet the deadline.
  If you have comments, please email me at martin.johnsson@slu.se}.
Because whole-genome sequence data is much more expensive than the SNP
chip genotypes currently used, it needs to deliver a large and
consistent improvement to be worthwhile.

The major achievement of livestock genomics in the past few decades was
the implementation of genomic selection. After mixed results with
marker-assisted selection --- indisputable successes with damaging
alleles of large effect (Knol et al., 2016; Schütz et al., 2008), the
detection of and selection against which have now become fairly routine
(Georges et al., 2019), and questionable usefulness for complex traits
(Dekkers, 2004; Lowe and Bruce, 2019) --- the combination of SNP
genotyping chips that cover the whole genome in markers, and estimation
methods that surmounted the \(p \gg n\) problem of simultaneously
dealing with many markers, made genomic selection possible. Nowadays,
large breeding programs are likely to have more genotyped animals than
markers, but treating marker effects as random still makes conceptual
sense.

Genomic selection has deep roots, going back at least to discussions
about selection on single loci (Fernando and Grossman, 1989; Smith,
1967; Soller, 1978), but at some point in the late 1990s, the field
shifted its focus from identifying key loci to use in marker-assisted
selection to treating the whole genome statistically (Haley and
Visscher, 1998; Lande and Thompson, 1990; Meuwissen et al., 2001;
Nejati-Javaremi et al., 1997). Implementation happened first in dairy
cattle breeding (Wiggans et al., 2017), later in pigs (Knol et al.,
2016), poultry (Wolc et al., 2016), and many other animal and plant
breeding programs (Hickey et al., 2017).

Thanks to its role in enabling genomic selection, the SNP chip, i.e., a
family of high-throughput array-based methods for SNP genotyping
(reviewed by Ragoussis 2009), is in the running for the title of most
impactful genomic technology. The SNP chip has attractive properties:
enough markers for genome-wide genotyping, cheap and accurate, and gives
rise to well-behaved tabular data -- as opposed to sequence data, which
requires more computation, and raises questions about how to represent
the genetic information. Many routine analyses are built around SNP chip
data. With some linear algebra, SNP chip genotypes can be turned into a
similarity matrix (i.e., genomic relationship matrix) that can be
plugged in as a variance---covariance matrix in a linear mixed model
(VanRaden, 2008). That is the essence of genomic selection. There is a
whole technical literature on how these models can be fitted
efficiently, evaluated and incorporate as much data as possible
(reviewed by Misztal et al., 2020).

\hypertarget{the-current-state-of-genomic-prediction-with-whole-genome-sequencing}{%
\section{The current state of genomic prediction with whole-genome
sequencing}\label{the-current-state-of-genomic-prediction-with-whole-genome-sequencing}}

Replacing SNP chip genotyping with whole-genome sequencing seemed like
an attractive next step for genomic prediction. While sequencing is much
more expensive, it has several purported benefits for genomic selection.
Meuwissen \& Goddard (2010) simulated genomic prediction with sequence
data and concluded that it would improve accuracy, and could
``revolutionize genomic selection in livestock''. The most natural
improvement to imagine is better accuracy of selection, but one might
also hope for better persistence of accuracy over subsequent
generations, and generalizability between populations (Hickey, 2013):

\begin{quote}
\emph{GS2.0 is a label that could be given to the type of GS that will
emerge in the next 5~years. \ldots{} potentially, millions of animals
will have data obtained by sequencing. If this is the case, GS2.0 will
accumulate the information required for utilizing both linkage
disequilibrium and causative nucleotides when making predictions about
breeding value. \ldots{} This will increase the accuracy and persistency
of predictions, could rescue the promise of across breed prediction and
make the explicit use of the millions of} de-novo \emph{mutations that
arise naturally in our breeding populations possible.}
\end{quote}

Compared to a SNP chip that can only type the genetic variants it was
designed to type, sequencing finds more variants, therefore has less
ascertainment bias, and has the potential to genotype more causative
variants. The typical SNP chip for farm animal might contain some 50,000
variants, whereas short read whole-genome sequencing routinely lets you
detect millions. The typical SNP chip will type common variants
ascertained in particular populations, whereas whole-genome sequencing
will detect variants in a less biased fashion (Geibel et al., 2021),
albeit not completely without reference genome bias (Ros-Freixedes et
al., 2018). Therefore, you would have to be very lucky for a typical SNP
chip to directly genotype causative variants (except known large-effect
variants when it has been designed to do so, e.g., Mullen et al.,
(2013)); sequence data, however, may have a chance to genotype the
causative variant directly. Finally, sequence data may be able to detect
other types of variants than single nucleotide variants, at least some
of the time.

Despite this appeal, both simulations and empirical results suggest that
genomic selection with sequence data does not yet work particularly
well. Using millions of variants from whole-genome sequencing, in
combination with imputation, is often no more accurate or even less
accurate than a SNP chip. Several studies (Moghaddar et al., 2019;
Raymond et al., 2018a; van Binsbergen et al., 2015; van den Berg et al.,
2017; VanRaden et al., 2017) found little to no benefit to using full
whole-genome sequence data for genomic prediction --- that is, not
pre-selecting any subset of variants, but using the millions of variants
directly. Raymond et al. (2018a), who found several cases where sequence
data \emph{decreased} the accuracy compared to SNP chip data, called it
a ``dilution effect'', where the many non-causal variants hampered
estimation. This is consistent with previous results that show little
improvement from increasing SNP chip marker density (Erbe et al., 2012;
Ilska, 2015).

The better method appears to be to use sequence data to pre-select a
subset of variants enriched for associations with traits and use them
for prediction, either as a bespoke ``in silico SNP chip'' or as a
supplement to an established SNP chip. However, even with this method,
benefits are relatively small and inconsistent between traits,
populations and methods. For example, Brøndum et al., (2015) found that
adding some 1600 markers selected from genome-wide association with
imputed sequence data to the 54k SNP chip improved accuracy for by a few
percentage points. Similarly, VanRaden et al. (2017) used imputed
whole-genome sequence data to select single nucleotide variants and add
them to the 60k set of SNPs used routinely; this led to improvement for
most traits, by 2.7 percentage points of reliability (i.e., the square
accuracy) on average. Moghaddar et al., (2019) used imputed-whole genome
sequence data from sheep to select SNPs and add them to a 50k SNP chip;
this led to increases in accuracy for most traits, on average 8-10
percentage points for different populations and methods. On the other
hand, Veerkamp et al. (Veerkamp et al., 2016) found no benefit from
pre-selected variants, and neither did Calus et al. (Calus et al., 2016)
when analysing the same data with a more sophisticated method. In pigs,
(Ros-Freixedes et al., 2022a) found inconsistent benefits between lines
and traits, but an average increase of 2.5 percentage points of
accuracy.

The situation is similar in multi-breed prediction scenarios. Despite
the idea that whole-genome sequence data might overcome the difference
in linkage disequilibrium between populations and improve across-breed
prediction, the accuracy gains from sequencing are small and
inconsistent. Several attempts have found small improvement to
prediction between breeds or genetic lines with pre-selected markers
from whole-genome sequence data (Meuwissen et al., 2021; Raymond et al.,
2018a, 2018b; Ros-Freixedes et al., 2022a; van den Berg et al., 2017).
For example, Raymond et al. (2018b) and Meuwissen et al. (2021) both
found minor increases from whole-genome sequence in multi-breed
scenarios where a small breed was supplemented with data from bigger
breeds. They both used methods that put higher weight on strongly
associated markers, using pre-selection and a separate relationship
matrix or a Bayesian variable selection method, respectively.
Ros-Freixedes et al., (2022a) found that with whole-genome sequence,
multi-line prediction was systematically worse than single-line
prediction, but compared to multi-line prediction with the SNP chip, the
relative improvement was greater. Thus, there is some truth to the idea
that multi-breed prediction benefits more from sequence data than
within-breed prediction, but the benefits are small and inconsistent.

Genomic prediction with sequence variants does not even perform that
well in simulations. The early simulations optimistically promised
substantially higher accuracy from whole genome sequence than SNP chips,
an increase in accuracy with marker density, and an additional increase
from being able to genotype the causative variants (Meuwissen and
Goddard, 2010). However, Meuwissen and Goddard already noted that a more
realistic population structure with more extensive linkage
disequilibrium would make improvement from sequence data less dramatic
than the one they simulated. Subsequently, MacLeod et al. (2014) found
very little benefit from sequence data over SNP chip data when the
population history was simulated to be similar to the history of cattle,
with an effectively small population and a historical population decline
due to domestication and breed formation. These results suggest that
there is little benefit to be gained from sequence data even when the
causative variants are included. Clark et al. (2011) found only a
relatively small difference between sequence data and a mid-density SNP
chip, especially when there were many causative variants.

Fragomeni et al. (2017) simulated the contemporary strategy of
pre-selecting causative variants to add to SNP chips. Even when all the
true causative variants were included, that only led to a modest
increase in accuracy. For the strategy to bring big benefits, they
needed not only the identity, but also the true effect size of each
variant, in order to be able to weight the causative variants
appropriately. They were unsuccessful in estimating these effects
accurately from genome-wide association studies, presumably due to
linkage disequilibrium. Jang et al. (Jang et al., 2023) simulated the
process of pre-selection by genome-wide association studies, exploring
under which conditions large effects can be identified to supplement the
SNP chip. They concluded that:

\begin{quote}
\emph{Even when variants are accurately identified, their inclusion in
prediction models has limited benefits.}
\end{quote}

Perez-Encisco et al. (2015) also found little improvement from
whole-genome sequence data and little improvement from pre-selection of
variants based on genome-wide association. However, their model assumed
that the simulated causative variants were located in a particular
subset of causative genes, and if those causative genes could be
identified accurately enough, they can be used as prior information to
give higher weight to variants. That means that their results support a
strategy of weighting variants based on biological priors (such as based
on functional genomics data), if that prior information can accurately
enrich for causative variants (in this model: by detecting causative
genes). We will return to this strategy below.

In summary, the hope that genomic selection with whole-genome sequencing
will allow accurate tracking of causative variants to give rise to
highly accurate and persistent genomic prediction, that works across
time and populations, is yet to be achieved. Whereas sequence data may
sometimes improve genomic selection accuracy, it is by no means a
game-changer similar to the introduction of genomic selection with SNP
chips. For the most part, this paper will take the position that this
lack of improvement form whole-genome sequence data is disappointing,
and a problem to be solved or at least explained. However, a positive
outlook is also possible. In some ways, it is good news that genomic
prediction with SNP chips is doing so well compared to the more
expensive and cumbersome sequence data.

There is a developing theoretical literature that attempts to explain
this limited success of genomic prediction with sequence data. The idea
is that effectively small populations, such as farm animal populations,
contain little enough genomic variation, that the bulk of this variation
can be captured with a typical SNP chip. We can think of the genome of a
population as a collection of genomic segments, that is, pieces of DNA
carrying unique combinations of variants. To track the genomic
variation, we only need enough markers that we track most of the
segments. This means that genomic selection with SNP chips uniformly
spaced along the genome will work well, and that it is hard to improve
upon by adding markers. To a first approximation, putting two markers on
the same segment adds nothing but estimation problems. Even if we
\emph{do} genotype the causative variant, the causative variant will be
confounded with everything else on the same segment.

\hypertarget{mental-models-of-genomic-selection}{%
\section{Mental models of genomic
selection}\label{mental-models-of-genomic-selection}}

In this section, I will survey such models of genomic selection with an
eye toward understanding the lack of success with sequence data. I will
concentrate on verbal models that guide intuition, but in each case the
authors also present formal models in the form of equations, simulations
or both.

The image sketched above of the population as a collection of segments
carrying a causative variant, or not, and a marker, or not, comes with a
model of genomic prediction accuracy presented by Goddard (2009). To
capture the fact that linkage puts a limit on how many markers are
needed to cover the genome, there is perfect linkage disequilibrium
within segment, and none between segments. Based on models from Sved
(1971) and Stam (1980), he derived formulas for the expected number of
segments in an ideal population, and the probability that two variants
fall on the same segment. The reciprocal of that probability is the
effective number of segments (loci) \emph{M\textsubscript{e}}. It is as
if the genome consisted of \emph{M\textsubscript{e}} little chromosomes,
each without recombination on them. Alternatively, this number can be
thought of from the perspective of realised genomic relationship between
individuals in a population (Goddard, 2009; Goddard et al., 2011). Real
populations can be numerically matched to ideal populations based on
their variance of relationship, like how we assign effective population
sizes to real populations based on rate of inbreeding or variance of
allele frequency.

While these formulas do not work great for predicting genomic selection
accuracy in practice (Brard and Ricard, 2015), the model is a starting
point for thinking about how genomic selection works. In particular, it
leads to two conclusions about genomes in populations: First, there is a
limit to the number of markers needed to track segments. Second, there
is a limit to the granularity of causative variants. Even if there are
more than one causative variant on a segment, from a statistical
perspective, that only modifies the net effect of the segment, but until
they are separated by recombination, they effectively work as one
causative variant. However, other research suggests that tight linkage
disequilibrium on short segments is not necessarily the most important
mechanism of genomic selection.

Habier et al. (2013, 2007) designed simulation scenarios ---
manipulating relatedness between training and testing set and placing
causative variants on the same or on different chromosomes --- in order
to separate different potential sources of genomic selection accuracy.
The first study (Habier et al., 2007) demonstrated genomic prediction
even in the absence of tight linkage disequilibrium between causative
variants and markers on segments. That is, when causative variants were
placed on different chromosomes than markers, genomic selection could
still work on the relatedness between individuals. In the second study
(Habier et al., 2013), they created scenarios to quantify the
contribution of linkage disequilibrium in founders, cosegregation within
families, and relationship between families. They found that most of the
genomic prediction accuracy derived from linkage disequilibrium in the
founder population, which in this simulation was one generation back, as
the pedigree was a set of half-sib families. Taking a different
simulation strategy, Wientjes et al. (Wientjes et al., 2013) generated
synthetic selection candidates --- either based on allele frequencies,
linkage disequilibrium, haplotype segments, or whole chromosomes ---
compared to real genotypes drawn from the reference population. They
evaluated the expected accuracy by predicting it from equations in the
case of synthetic genotypes, and by cross-validation in the case of real
individuals. The accuracy with real individuals was much higher than
with any of the synthetic scenarios, and since the one feature the real
individuals have that the synthetic genotypes lack is close relationship
to the reference population, they concluded that close relationship is
the most important driver of genomic prediction accuracy. These studies
come to quantitatively different conclusions about the drivers of
accuracy, but they both illustrate the limitations of thinking of a
population under genomic selection as a collection of independent
segments.

Pocrnic et al. (Pocrnic et al., 2019) presented a competing verbal
model, describing genomic selection as based on clusters of segments
(referred to in the paper as ``clusters of independent chromosome
segments'', ``clusters of haplotypes'', and ``clusters of
\emph{M\textsubscript{e}}''), rather than independent segments. In a
population, variants are quantitatively associated, and there are some
major axes of variation that can be found among the genotypes. The
accuracy of genomic selection, they propose, is driven by tracking the
most important collections of segments that are currently inherited
together.

The formal model that goes with this idea is an eigendecomposition of
the genomic relationship matrix (Pocrnic et al., 2019, 2016a, 2016b).
They created reduced matrices that only included information from the
top eigenvectors of the full genomic relationship matrix, and tested
their performance for prediction. When enough eigenvalues were included,
a reduced matrix was able to produce effectively the same prediction
accuracy of the full one. It turns out that the dimensionality of the
genotype matrix of a typical farm animal population is quite limited.
The clusters with the largest eigenvalues contribute the most to
accuracy, so that prediction works relatively well even with a small
amount of genetic information, and then increases only slightly when
smaller clusters are added (Pocrnic et al., 2019). Misztal et al. (2022)
repeated this interpretation as an explanation for why multi-breed
genomic prediction is difficult, and not much improved by whole-genome
sequence data either.

There are several questions to ask about these clusters: How do they
relate to other descriptions of genetic structure, such as haplotype
blocks and linkage disequilibrium heatmaps? To what extent do they
reflect local haplotypes on a chromosome, or span different chromosomes?
How do they relate to within and between chromosome genetic covariances,
and how do they relate to Habier et al.'s and Wientjes et al.'s sources
of linkage disequilibrium? How do they change with selection? How does
that relate to decay of genomic prediction accuracy over generations?

At any rate, both in the independent segment model and the cluster model
there is a limit to the genetic information contained in your sample ---
due to the number of animals and marker density, but also an intrinsic
limit to the granularity of genetic information, that in some ways come
down to the structure of the genome and the effective size of the
population. Because farm animal populations are small, at some point, it
does not matter much how many genetic variants we genotype, because they
contain more of the same information, for a given set of animals.
Sampling more individuals at the same time as increasing marker density
would reveal more information, albeit at diminishing returns. It seems
to me that this limit was reached earlier than geneticists expected --
or, alternatively, problems with estimation and representation prevent
our models from making use of the additional information from many more
genetic variants.

This also means that the marker effects estimated in genomic selection,
even when sequence data allows (near) complete genotyping of all
variants, are ephemeral because they reflect the net effect of genomic
segments, or clusters of genomic segments, rather than the isolated
effect of individual causative variants. There are two parts to this
problem. On the one hand, it is hard to accurately estimate the effect
of sequence variants because they are allelically associated. On the
other hand, when genomic breeding values are formed, the genotypes and
estimated variant effects are multiplied and summed together again. Even
if the effects are estimated in a way that more accurately resolves
causative variants, we could arrive at an equally accurate breeding
value by assigning the effects to noncausal but associated variants,
because predictions of breeding values are linear combinations of those
effects. Accordingly, whole-genome sequence may be more valuable for
fine-mapping of variants than it is for prediction.

Another limitation to our knowledge of genetic effects comes about
because our estimates represent not only the net effect of all causative
variants in linkage disequilibrium, but the net additive effect when
averaging over any genetic interactions they participate in. That is,
marker effects are linear coefficients of trait values on variant
dosages. In the presence of non-additive effects, those linear
coefficients might still provide a decent estimate, but they are liable
to change as the allele frequencies at the variant itself and its
interaction partners change (like traditional average effects of alleles
(Falconer and Mackay, 1996, pp. 112--119)). Legarra et al. (2021)
derived equations for the change in additive effects between populations
and generations, by taking derivatives of the statistical effects with
respect to allele frequency, then using Taylor expansion to create an
approximation of the change around the allele frequency in a focal
population. The model illustrates that there are dual reasons why
genomic prediction accuracy decreases with genetic distance: not only
because the associations between variants change, but also because
allele frequency differences at interacting causative variants change
the net effect of the variants on traits.

\hypertarget{sequence-data-how-to-make-them-pay}{%
\section{Sequence data: how to make them
pay}\label{sequence-data-how-to-make-them-pay}}

Therefore, the biggest future challenge for livestock genomics, as I see
it, is to get value for the money and work that goes into sequence data,
in the form of improvements to breeding practice. To do this, we need to
overcome the low dimensionality of the genetic information --- or the
small number of effective segments --- with some clever strategy. I
speculate that there are three main attacks on this problem.

\hypertarget{better-modelling-of-genomic-segments}{%
\subsection{Better modelling of genomic
segments}\label{better-modelling-of-genomic-segments}}

First, perhaps we could improve the way we detect and represent genomic
segments from sequence data. There are two parts to this: one is about
improving inference of the genotypes with imputation, sequencing
strategies, etc, and the other is about improving the representation of
the genomes, and making explicit use of the information from segmental
structure of the data.

One weakness of previous research is that it, universally, used imputed
sequence data. This limitation is unavoidable, unless a technological
breakthrough makes whole-genome sequencing as cheap as SNP chip
genotyping. Imputation from sequence data is still not as
straightforward as imputation of SNP chip data, but there are now
several strategies based on Hidden Markov models (Browning et al., 2021,
2018; Delaneau et al., 2019) or multilocus segregation analysis
(Ros-Freixedes et al., 2020; Whalen et al., 2018). Any imputation to
sequence-level will rely on some less dense set of markers to impute
from, derived from extremely low-coverage sequencing, reduced
representation sequencing, or SNP chip genotyping. While modern
imputation methods perform well, and can be checked on held-out data,
imputed data always has some limitations, like the ability to recover
rare variants and resolve the location of new recombination events.

More sequence data might be needed, especially to capture more of the
rare variants. Recent results (Ros-Freixedes et al., 2022a) suggest that
larger sample sizes might help, as whole-genome sequence data tended to
do better in populations where the training sets were larger. More
sequenced individuals mean more individuals with high density genotypes
and more ability to detect rare variants. There are many rare and
population-specific variants that may contribute to traits
(Ros-Freixedes et al., 2022b), and simulations that vary the allele
frequency spectrum show that rare causative variants make genomic
prediction more difficult (MacLeod et al., 2014; Wientjes et al., 2015).
Part of the solution may be to sequence more, which adds to the cost. If
very low coverage sequencing could become an alternative to SNP chip
genotyping, as some have suggested (Snelling et al., 2020), that might
help contribute sequence information. However, we should keep in mind
that simulations that have perfect data still struggle with genomic
prediction with whole-genome sequence, suggesting that even if
imputation accuracy were perfect, there would be additional issues.

It might also be possible to find new representations of
population-scale whole-genome sequence data that facilitate genomic
prediction. Currently, the options are either to put all variants into a
large, potentially millions-by-millions, matrix and letting a model sort
them out --- a modelling strategy that is not at all successful --- or
to use a pre-selection method to find a smaller set of more relevant
markers, either by literal pre-selection that subsets the variants that
the model is seeing or by some model that does variable selection based
on data. A third option might be to find a representation of genome
segments that capture the relevant structure, with the ambition to fit a
model that does not struggle so much when given millions of variants.

At least parts of the long-standing line of research on haplotype models
fall in this category. The intuition is that because haplotype models
account for the associations of variants close together in the genome,
they are more realistic than models that treat markers independently.
Haplotype models have been tried many times, usually on SNP chip data,
with variable benefits. Haplotype models come with practical problems of
defining haplotypes. In recombining regions of a genome, segments may
start at any point in a given individual, creating fuzzy borders between
haplotypes. We need some methods to create windows or blocks, that are
often arbitrary. Proposals to better deal with this includes defining
windows based on recombination hotspots (Oppong et al., 2022), haplotype
block methods that create overlapping segments (Pook et al., 2019), and
haplotype clustering methods (Browning and Browning, 2007).

Furthermore, because of the many combinations of alleles within a
window, there are likely to be many haplotypes, especially if applied to
sequence data. This often means that the problem of fitting many
variants with two alleles turns into the problem of fitting a smaller
number of windows with more alleles. To solve this problem, one must
find representations of relationships between haplotypes. Several
attempts have been made using similarities between haplotypes (Hickey et
al., 2013), grouping consecutive markers based on linkage disequilibrium
(Cuyabano et al., 2015, 2014), local convolutional neural networks that
represent regions of the genome as part of neural network structure
(Pook et al., 2020), and by phylogenetic analysis of haplotypes (Edriss
et al., 2013; Selle et al., 2021). Selle et al. (2021), who developed a
model for prediction based on phylogenetic relationships between
non-recombining haplotypes, propose that recently developed methods for
inferring genealogy along the genome in the presence of recombination
and representing it as so-called tree sequences (Kelleher et al., 2019)
may be useful.

Non-SNP variants present further complications for our representations
of the genome. Genomic prediction models can easily represent genotypes
at biallelic non-overlapping variants. Each variant corresponds to one
column of the genotype matrix. Any variant set involving non-SNPs,
however, may contain overlaps. Take the simple example of a SNP that
overlaps an indel. Representing a biallelic indel is just as easy as a
biallelic SNP, but if they overlap there may be chromosomes that have a
null allele of the SNP because they carry the allele that deletes the
region around the SNP. In the Variant Call Format used for short-read
sequence results, this is represented by the asterisk `*' allele. This
situation is similar to the haplotype models, where we end up with
multi-allelic variants, and judgement calls about what variants to group
and not. The full sequence variation is messier than a grid of SNPs, and
harder to represent neatly.

\hypertarget{inclusion-of-undetected-genetic-variation}{%
\subsection{Inclusion of undetected genetic
variation}\label{inclusion-of-undetected-genetic-variation}}

Proposals for whole-genome sequence data for genomic prediction usually
emphasise that sequence data can directly genotype the causative
variants (Hickey, 2013; Meuwissen and Goddard, 2010). However, many
types of variants are likely to be absent from the imputed whole-genome
sequence data that has been used so far, which is based on short-read
sequencing and reference-guided analysis. Therefore, more complete
detection of variants is another avenue for improvement.

Most imputed sequence datasets are limited to single nucleotide variants
and short insertions/deletions. Whole-genome sequencing is good at
detecting single nucleotide variants, routinely finding millions of them
in small population samples. However, in repetitive regions of the
genome, even single nucleotide variants and short insertions/deletions
are hard to detect. Variants from short read sequencing are routinely
filtered by different sets of (heuristic, ad hoc) filters, including
proximity filtering, excluding multiallelic sites or exclusion of
repetitive sequence (Daetwyler et al., 2014; Van der Auwera et al.,
2013). These filters are evaluated (if at all) by comparing the results
to previous datasets or by expected population genetic properties of the
variants detected (e.g., transversion/transition ratio). This certainly
improves the quality of the variants that are detected, but also means
that no dataset can realistically claim to be a complete compendium even
of the common single nucleotide variants and short insertions/deletions
in a population.

Larger-scale structural variants are even harder to detect without
long-read sequencing or even genome assembly (Nguyen et al., 2023).
Currently, these methods are prohibitively expensive for
population-scale analyses. However, there are methods for genotyping
structural variants from short read sequence once they are known (Ebler
et al., 2022; Hickey et al., 2020), suggesting that it might be possible
to sequence a smaller number of animals and impute the structural
variants. The fluorescence intensity signal from SNP chips used for
genotyping also contain some information about copy number, which might
also contribute.

Because accurate structural variant detection requires population-level
long read sequencing, and research has concentrated on between-breed
comparisons, it is not clear how much structural variability there is in
livestock populations, but likely a lot. We can get an idea from studies
with short read sequencing and copy number analysis of SNP chips. Butty
et al. (2020) combined short read sequencing and SNP chip copy number
detection to identify a high-confidence set of structural variants that
covered a total of 7.5 Mbp (0.3\% of reference genome length) in
Holstein cattle. Chen et al. (2021) detected structural variants from
short read sequencing and imputed them to SNP chip genotyped cattle.
They detected 20 Mbp of structurally variable sequence within Holstein
cattle (0.7\% of reference genome length), and 3.5 Mbp of structurally
variable sequence within Jersey cattle (0.1\% of reference genome
length). Imputed structural variants explained up to 8\% of the genetic
variance in milk traits, fertility and conformation, and did not
appreciably increase genomic prediction accuracy. These numbers are
likely to be underestimates because long read assembly-based analysis in
humans discovered more than three times as much structural variation as
short read sequencing (Ebert et al., 2021). Two large single nucleotide
variant datasets of cattle (Hayes and Daetwyler, 2019) and pigs
(Ros-Freixedes et al., 2022b) contain 43 million and 39 million SNPs,
respectively, corresponding to 1.6\% of cattle genome length and 1.5\%
of pig reference genome length. Therefore, it seems likely that farm
animals (like humans) have more basepairs affected by structural
variation than by single nucleotide variants.

Unfortunately, a discouragingly high number of damaging variants and
causative variants for monogenic traits and pigmentation that are known
(or strongly suspected) have turned out to be caused by structural
variants (Dorshorst et al., 2011; Durkin et al., 2012; Gunnarsson et
al., 2011; Imsland et al., 2012; Kadri et al., 2014; Mishra et al.,
2017; Rubin et al., 2012; Wang et al., 2013; Wiedemar et al., 2014;
Wright et al., 2009). Also, genetic mapping of gene expression (eQTL
mapping) studies in humans that include structural variants have found
an enrichment of structural variants among the most strongly associated
variants (Chiang et al., 2017; Ebert et al., 2021; Sudmant et al.,
2015), and found larger effects associated with structural variants than
single nucleotide variants. This is tentative evidence that structural
variants are particularly likely to be causative, and bad news if we
hoped that variants called from short-read sequencing would include
causative variants.

Some of this structural variation is going to be tagged by linkage
disequilibrium with surrounding single nucleotide variants, but not all.
Yan et al. (2021) found that, in humans, about half of their structural
variants were in what they term strong to moderate (r\textsuperscript{2}
\textgreater{} 0.5) association with at least one surrounding variant.
In the chicken, Geibel et al. (2022) found that deletions were well
tagged by single nucleotide variants, with linkage disequilibrium
comparable to the values between pairs of single nucleotide variants,
but that other structural variants types (duplications, inversions, and
translocations) were poorly tagged. In this case, however, the
structural variants were called with short read sequencing and
relatively low coverage, that likely has low accuracy for duplications,
inversions and translocations. Similarly, Xu et al. (2014) detected copy
number variants from fluorescence intensity on SNP chips, and found a
subset of copy number variants that were well tagged by surrounding
single nucleotide variants, and another subset that was not.

Apart from the representation issues (see previous section), structural
variants are likely to have different mutation rate distributions than
SNPs and can affect local recombination rate. For example, structural
variation often happens in already repetitive regions, with biases
towards similar sequences. Gene conversion may also play a role, in
particular in highly repetitive regions. These population genetic
differences from SNPs are likely to affect the association patterns with
surrounding variants (reviewed by (Conrad and Hurles, 2007)). The most
important undetected variants to include would be the ones that are
poorly tagged by variants already typed. They are therefore also more
likely to be hard to impute correctly, and less likely to be
pre-selected based on genome-wide association -- since both imputation
and genome-wide association rely on allelic association.

Part of the problem is that structural variants tend to occur in
repetitive regions of the genome that are hard to sequence and genotype,
and another part is that structural variants can interfere with the
genotyping of neighbouring variants, by changing flanking sequence,
changing the order of the genetic map, causing null alleles or
artificial heterozygotes by duplication and so on. Thus, the low linkage
disequilibrium around many structural variants may be partially due to
biology and partially due to technical difficulties (Geibel et al.,
2022; Yan et al., 2021).

In summary, there are good reasons to expect that structural variants
will be common and that many causative variants will be structural.
Whether structural variant detection can be used in genomic prediction
will depend on the ability to genotype them at scale. However, again,
the simulation studies above used perfect data with inclusion of the
causative variants (together with all other sequence variants in MacLeod
et al. (2014); together with non-causal SNP chip markers in Fragomeni et
al. (2017)). While they did not explicitly model structural variants
(but rather abstract biallelic causative variants), the result that
genomic prediction with sequence data does not work well even when all
simulated causative variants are genotyped suggests that detection of
causative variants is not the main obstacle.

Another cause of undetected variation might be missing regions from
reference assemblies, either because of hard-to-assemble regions or
because genetic differences between the reference assemblies and the
animals of interest. Several projects try to remedy these omissions by
creating new genome assemblies from divergent breeds and aggregating
them into pan-genomes that aim to represent the whole gene pool of a
species. This amounts to assuming that there are enough undetected
sequences in the genome that are important to traits but uncorrelated
with typed segments. A recent pan-genome effort including four cattle
breeds (three European \emph{Bos taurus} \emph{taurus} breeds and
Nellore, which is \emph{Bos taurus indicus}) as well as another species
(gaur, \emph{Bos gaurus}) added 82.5 Mbp sequence not in the reference
genome (Leonard et al., 2022), about 3\% of the reference genome length,
whereas a pangenome of three European cattle breeds and two African
cattle breeds (the \emph{taurus} N'Dama and \emph{taurus x indicus}
cross Sanga Ankole) added 116 Mbp, about 4\% of the reference genome
length (Talenti et al., 2022) --- or 20.5 Mbp, filtered down to what
they term high-quality non-repetitive sequence, that is 0.7\% of the
reference genome length. Similarly, a pig pangenome based on five
European and six Chinese breeds added 72.5 Mbp, about 3\% of the pig
reference genome length (Tian et al., 2019). This suggests that a pig or
cattle breed may contain up to a few percent of sequence not included in
the reference genome. Whether better tracking that sequence through a
breed-specific assembly will improve genomic prediction will depend on
whether those breed-specific sequences are enriched for genetic variance
in important traits, and to what extent they are already tagged by
marker panels used for genomic prediction through linkage
disequilibrium.

\hypertarget{use-of-functional-genomic-information}{%
\subsection{Use of functional genomic
information}\label{use-of-functional-genomic-information}}

The first two avenues for improvement in genomic prediction with
sequence data were about detecting and describing the genetic variation
within populations, whereas the third is about adding functional rather
than purely genetic information. A large amount of chromatin and gene
expression data has started to accumulate from projects within the FAANG
collaboration, and similar efforts (Giuffra et al., 2019; Halstead et
al., 2020; Kern et al., 2021; Salavati et al., 2022; Zhao et al., 2021).

There is some evidence that functional genomic data may help enrich for
variance in quantitative traits. For example, Wang et al. (2017) found
that putative enhancers identified by chromatin immunoprecipitation
sequencing of histone marks were enriched for genetic variation in milk
production traits in cattle. In a series of studies, Xiang et al. (2021,
2019) created a score for prioritising variants for pre-selection, that
included functional genomic data in combination with evolutionary
conservation scores and quantitative genetic analyses, which was used to
create a custom SNP chip with improved prediction accuracy. The FarmGTEx
collaboration has created mega-analyses of the genetic basis of gene
expression by pooling RNA-sequencing data from many studies and imputing
genotypes from the reads (Liu et al., 2022; The FarmGTEx-PigGTEx
Consortium et al., 2022). Combining this type of data with genomic
prediction, Xiang et al. (2022b) found that variants associated with
gene expression are enriched for genetic variation in selected traits,
to the point where 70\% of the variance can be accounted for by a set of
850,000 variants, which is more than a size-matched random selection. It
remains to be seen what it translates to in terms of genomic prediction
accuracy when such methods are tested at the scale of a breeding
program.

The idea is to use functional genomic data to prioritise for
pre-selection or put extra weight on such variants that have supporting
molecular evidence. There are ambitious proposals on how to layer other
kinds of data on top of each other --- from the open chromatin and gene
expression that are available today to functional assays that can be
performed at scale such as CRISPR inhibition/activation screens or
massively parallel reporter assays. They all potentially give
genome-wide information about variant function that is, in some sense,
independent of trait variation and the constraints of linkage
disequilibrium and limited dimensionality.

The FAANG to fork paper (Clark et al. 2020) expresses this vision
clearly:

\begin{quote}
\emph{Most of these causal variants, with small effects, are likely to
be located in regulatory sequences and impact complex traits through
changes in gene expression ... Thus, it is expected that improvements in
prediction accuracy can be achieved by filtering the genetic marker
information based upon whether the genetic variants reside in functional
sequences and developing robust prediction models that can accommodate
the biological priors. ... As many more suitable datasets will become
available in the next 5 years, improving and adapting these methods to
enhance genomic prediction accuracy, whilst conserving genetic
diversity, across farmed animal species will be a priority for FAANG.}
\end{quote}

Expressed in terms of our mental models of genomic selection, proposals
to combine functional genomic data with genomic selection hypothesise
that functional genomic data, when summarised over many different assays
and tissues will yield information about causative variants is accurate
enough for a genomic prediction model to accurately estimate effects for
variants that are located on the same genomic segment. The
identification does not need to be definite, but accurate enough to
improve estimates of the variants' effects. The functional information
needs, as it were, to break ties between variants that are genetically
confounded.

For example, one might identify a relevant tissue where gene expression
traits, collectively, explain a substantial proportion of genetic
variance (such as the udder for lactation traits in cattle (Liu et al.,
2022; Prowse-Wilkins et al., 2022)). One might then use a massively
parallel reporter assay in a cell model of the udder to test all the
variants in active chromatin in the udder (as proposed by Littlejohn et
al. (2022)), and perform genomic prediction based on the variants that
show allelic differences in the reporter assay. If recent results from
humans are anything to go by, this would likely be thousands of variants
(Abell et al., 2022). Because the information about gene-regulatory
causality in the reporter assay is independent of genetic analysis and
not confounded by linkage disequilibrium between variants, it might, if
it is specific and accurate enough, allow the right variant among a set
of correlated variants to receive a higher estimated effect.

This entails a couple of assumptions about functional genomic data.
First, that they contain distinct enough information to tell apart
functional and non-functional variants that are located close together
in the genome. Functional genomic methods often produce correlated
features, and struggle, for example, to tell apart variants located in
the same chromatin element (Liu et al., 2019). However, this correlation
is likely to range over a shorter scale than the extent of linkage
disequilibrium in farm animals, so it is likely to be an improvement. In
this respect, engineering-based methods like massively parallel reporter
assays might have an edge over observational methods like open chromatin
analyses. Methods that give high resolution about protein binding, such
as DNAse I hypersensitivity profiling might also help. Second, we have
to assume that functional genomics data contain specific enough
information that we can distinguish the causative variants that are
relevant to our trait of interest, when there are multiple genuine
causative variants for different traits. There are likely to be multiple
linked causative variants (Abell et al., 2022; Xiang et al., 2022a) for
many traits, and consequently a very large number of variants that are
genuinely causal for different traits will occur close to each other.
Here, methods that identify tissues and conditions that are enriched for
variance in particular traits (Liu et al., 2022) may be helpful to find
relevant tissue-specific variant annotations.

The simulations by Fragomeni et al. (2017) suggest that to derive the
full benefit from sequence variants, we would need not only to identify
them, but to estimate their effects in order to weight them properly in
the genomic prediction model. When weights were estimated by genome-wide
association, both in their simulation and later work by Jang et al.
(2022), however, there was little benefit to weighting. If estimation of
effect sizes is needed, that would be an additional problem, because
functional genomics analyses are usually concerned with finding the
identity of the variants and there is little attention to estimating
their effect on downstream traits. Because the effect of genetic
variants depends on complex, non-linear, and largely unknown gene
regulatory and physiological systems, it is not clear how to translate
functional genomic effects (such as the fold change in chromatin
immunoprecipitation signal or transcript abundance) into effect sizes at
the trait level.

In a sense, functional genomic data in genomic prediction may be about
excluding irrelevant variants as much as it is about finding the
causative ones. If 70\% of the genetic variation in several quantitative
traits can be captured by variants associated with gene expression
(Xiang et al., 2022b), that suggests that a decent fraction of the
genome does not need to be accounted for in genomic prediction.
Similarly, Yengo et al. (2022) found that for the extremely polygenic
trait of human height, associated regions covering about 20\% of the
genome explained about half of the genetic variance. The fraction of the
genome that is associated would likely be greater in livestock due to
more extensive linkage disequilibrium, but the observation suggests
that, in principle, there is scope to cut down the search space.
However, the subset of the genome that matters might still be larger
than an ordinary SNP chip, and different for different traits.

\hypertarget{pay-for-what}{%
\subsection{Pay for what?}\label{pay-for-what}}

What kinds of information would be needed to know when sequencing is
worthwhile? The economy of using sequence data for genomic prediction
depends on what data already exists, what needs to be generated, and the
benefit to accuracy --- which unfortunately seems to be specific to
populations, traits, and methods.

The case looks favourable for using publicly available or already
generated sequence data in combination with a modified SNP chip.
VanRaden et al. (2017) compared the potential economic value of the
increased selection accuracy that they achieved with the cost of
sequencing the bulls contributed by the US to the 1000 Bull Genomes
Project, and argued that the return on investment was high. They did not
factor in the cost of genotyping an additional 17,000 SNPs per animal,
but presumably a somewhat higher density SNP chip is not prohibitively
expensive for a large organisation, and maybe one can make space in a
custom SNP panel by taking out markers that are monomorphic or rare on
the target population. If the sequence data is available, for example
from a research project, this seems like a reasonable exercise that
someone running a breeding program could do to potentially improve their
genomic selection accuracy.

However, if sequence data is not available, the value of starting
sequencing is less clear. In the long run, the additional accuracy for
certain traits conferred by whole-genome sequence data might be enough
to justify this investment --- especially if it comes out of a research
budget. After the large initial cost for sequencing the reference
animals for imputing sequence data to a population, it appears to be
possible (Ros-Freixedes et al. 2020, figure 4) to keep imputation
accuracy up for additional generations without sequencing effort. The
sustained accuracy makes sense, because barring new mutation, the
population is just re-shuffling the same genomic segments. With clever
computational methods, imputation might even become relatively
computationally affordable, as Browning et al. (2018) suggested with
their ``one-penny imputed genome'' for humans. The case for long-read
sequencing is less compelling because the sequencing is much more
expensive, and it is not as clear how structural variants are to be
imputed or genotyped after detection.

New genome assemblies for different breeds and pangenomes are being
generated because of their scientific interest. Presumably what is
needed to make use of novel breed-specific sequence detected is to
identify markers located in them, and add them to updated SNP chips.
Breeding organisations that own particular populations and lines that
have not been sequenced in public projects might generate the assemblies
themselves, e.g. (Derks et al., 2022). This is a one-time investment
that could also help other kinds of genetic analysis.

Similarly, functional genomic data is expensive and technically
difficult to generate, but does not need to be generated at
production-scale. Open chromatin assays are usually run only on a
handful of samples per tissue or cell type and condition. Genetic
mapping of molecular traits (like gene expression in eQTL mapping) needs
a genotyped population sample, but it is usually on the order of
hundreds of samples rather than thousands per tissue. However, recent
mega-analyses of cattle (Liu et al., 2022) and pig (The FarmGTEx-PigGTEx
Consortium et al., 2022) expression datasets suggest that the number of
detected associations increase with sample size, suggesting that current
sample sizes have low power.

Finally, there are additional computing costs associated with sequence
data analysis that are not negligible, if sequence data is to be used
routinely. Just storage of sequence data takes up orders of magnitude
more space than SNP chip data, and sequence imputation will need to be
repeated regularly, as will likely the genome-wide association studies
used for pre-selection of variants. On the one hand, these are areas of
active research where improvements can be expected, but on the other
hand, sequence data analysis is a far cry from the relative convenience
and routine of handling SNP chip data.

In summary, if the benefits of genomic prediction with sequence data for
particular traits and populations are consistent over time, and if
genotyping is not so expensive --- for example by updating a SNP chip
that already needs updating anyways, or by persistently accurate
imputation from sequence data that already exists --- they would be
worthwhile. If, on the other hand, the inconsistent accuracy between
populations and traits translates into inconsistent performance over
time, or larger investments in sequencing are needed, then the benefits
would be questionable.

\hypertarget{other-ways-to-make-use-of-sequence-data}{%
\subsection{Other ways to make use of sequence
data}\label{other-ways-to-make-use-of-sequence-data}}

One might also argue that there are other uses of sequencing and
functional genomic data, such as microbiome sequencing, that I have
neglected. With these ``other omics'', the idea is to use
population-scale functional genomics or microbiome sequencing for
prediction. The same logic applies. If these data are supposed to bring
predictive benefits to animal breeding, they will have to pay for
themselves; and currently, they are viciously expensive. Other omics
serve as high-dimensional phenotypes as well as genomic information ---
as opposed to SNP chip genotypes that are just genotypes, a microbiome
sample or an epigenomic sample may also contain useful information about
the environment that may be useful for management or environmental
monitoring. Further, high-dimensional omic phenotypes might be useful
for predicting traits of animals that are difficult to measure, such as
feed efficiency. Whether such omic prediction is more affordable than
measuring the trait itself will obviously depend on the trade-off
between prediction accuracy and costs. Still, it is difficult to see how
the economy of on-farm use of other omics will work out within the
foreseeable future. If it is hard to convince a farmer pressed for money
to genotype their cows, it appears impossible to justify metagenome
sequencing.

One might argue that I have excluded the most important task of
livestock genomics: to identify causative variants that are directly
useful for genome editing or marker-assisted selection. For monogenic
traits, causative identification is a feasible (as discussed by (Georges
et al., 2019) and evidenced by the programme at any animal genetics
conference) and useful for veterinary medicine and management of defects
in breeding programs. These applications also have a more favourable
costs and benefits because, potentially, one needs to sequence only a
few cases and controls, rather than target the whole population, and
generate functional data from a candidate gene in some relevant tissue
--- a study more akin to traditional experimental biology.

\hypertarget{conclusions}{%
\section{Conclusions}\label{conclusions}}

Nothing is new under the sun. That identification of causative variants
is hard follows from classical quantitative genetic theory, and early
calculations on marker-assisted selection already suggested that the
benefits of selection on known genetic variants is limited to large
effects (Soller, 1978). However, one might hope that the confluence of
new data, new data analysis methods and new models might help us make
better use of large-scale genomic data in the future.

\hypertarget{funding}{%
\section{Funding}\label{funding}}

The author acknowledges the financial support from Formas---a Swedish
Research Council for Sustainable Development Dnr 2020-01637.

\hypertarget{conflict-of-interest-disclosure}{%
\section{Conflict of interest
disclosure}\label{conflict-of-interest-disclosure}}

I declare that I have no financial conflicts of interest in relation to
the contents of this article. However, the reader should keep in mind
that the text may reflect my scientific interests and biases.

\hypertarget{references}{%
\section{References}\label{references}}

Abell, N.S., DeGorter, M.K., Gloudemans, M.J., Greenwald, E., Smith,
K.S., He, Z., Montgomery, S.B., 2022. Multiple causal variants underlie
genetic associations in humans. Science 375, 1247--1254.
https://doi.org/10.1126/science.abj5117

Brard, S., Ricard, A., 2015. Is the use of formulae a reliable way to
predict the accuracy of genomic selection? Journal of Animal Breeding
and Genetics 132, 207--217. https://doi.org/10.1111/jbg.12123

Brøndum, R., Su, G., Janss, L., Sahana, G., Guldbrandtsen, B., Boichard,
D., Lund, M., 2015. Quantitative trait loci markers derived from whole
genome sequence data increases the reliability of genomic prediction.
Journal of dairy science 98, 4107--4116.

Browning, B.L., Browning, S.R., 2007. Efficient multilocus association
testing for whole genome association studies using localized haplotype
clustering. Genetic Epidemiology 31, 365--375.
https://doi.org/10.1002/gepi.20216

Browning, B.L., Tian, X., Zhou, Y., Browning, S.R., 2021. Fast two-stage
phasing of large-scale sequence data. The American Journal of Human
Genetics 108, 1880--1890. https://doi.org/10.1016/j.ajhg.2021.08.005

Browning, B.L., Zhou, Y., Browning, S.R., 2018. A One-Penny Imputed
Genome from Next-Generation Reference Panels. The American Journal of
Human Genetics 103, 338--348. https://doi.org/10.1016/j.ajhg.2018.07.015

Butty, A.M., Chud, T.C.S., Miglior, F., Schenkel, F.S., Kommadath, A.,
Krivushin, K., Grant, J.R., Häfliger, I.M., Drögemüller, C., Cánovas,
A., Stothard, P., Baes, C.F., 2020. High confidence copy number variants
identified in Holstein dairy cattle from whole genome sequence and
genotype array data. Sci Rep 10, 8044.
https://doi.org/10.1038/s41598-020-64680-3

Calus, M.P.L., Bouwman, A.C., Schrooten, C., Veerkamp, R.F., 2016.
Efficient genomic prediction based on whole-genome sequence data using
split-and-merge Bayesian variable selection. Genetics Selection
Evolution 48, 49. https://doi.org/10.1186/s12711-016-0225-x

Chen, L., Pryce, J.E., Hayes, B.J., Daetwyler, H.D., 2021. Investigating
the Effect of Imputed Structural Variants from Whole-Genome Sequence on
Genome-Wide Association and Genomic Prediction in Dairy Cattle. Animals
11, 541.

Chiang, C., Scott, A.J., Davis, J.R., Tsang, E.K., Li, X., Kim, Y.,
Hadzic, T., Damani, F.N., Ganel, L., Montgomery, S.B., Battle, A.,
Conrad, D.F., Hall, I.M., 2017. The impact of structural variation on
human gene expression. Nat Genet 49, 692--699.
https://doi.org/10.1038/ng.3834

Clark, S.A., Hickey, J.M., van der Werf, J.H., 2011. Different models of
genetic variation and their effect on genomic evaluation. Genetics
Selection Evolution 43, 18. https://doi.org/10.1186/1297-9686-43-18

Conrad, D.F., Hurles, M.E., 2007. The population genetics of structural
variation. Nat Genet 39, S30--S36. https://doi.org/10.1038/ng2042

Cuyabano, B.C., Su, G., Lund, M.S., 2015. Selection of haplotype
variables from a high-density marker map for genomic prediction. Genet
Sel Evol 47, 61. https://doi.org/10.1186/s12711-015-0143-3

Cuyabano, B.C., Su, G., Lund, M.S., 2014. Genomic prediction of genetic
merit using LD-based haplotypes in the Nordic Holstein population. BMC
Genomics 15, 1171. https://doi.org/10.1186/1471-2164-15-1171

Daetwyler, H.D., Capitan, A., Pausch, H., Stothard, P., Van Binsbergen,
R., Brøndum, R.F., Liao, X., Djari, A., Rodriguez, S.C., Grohs, C.,
2014. Whole-genome sequencing of 234 bulls facilitates mapping of
monogenic and complex traits in cattle. Nature genetics 46, 858--865.

Dekkers, J.C.M., 2004. Commercial application of marker- and
gene-assisted selection in livestock: Strategies and lessons1,2. Journal
of Animal Science 82, E313--E328.
https://doi.org/10.2527/2004.8213\_supplE313x

Delaneau, O., Zagury, J.-F., Robinson, M.R., Marchini, J.L.,
Dermitzakis, E.T., 2019. Accurate, scalable and integrative haplotype
estimation. Nat Commun 10, 5436.
https://doi.org/10.1038/s41467-019-13225-y

Derks, M.F.L., Boshove, A., Harlizius, B., Sell-Kubiak, E., Lopes, M.,
Grindflek, E., Knol, E., Groenen, M., Gjuvsland, A.B., 2022. A
pan-genome of commercial pig breeds. Presented at the World Congress of
Genetics Applied to Livestock Production 2022.

Dorshorst, B., Molin, A.-M., Rubin, C.-J., Johansson, A.M., Strömstedt,
L., Pham, M.-H., Chen, C.-F., Hallböök, F., Ashwell, C., Andersson, L.,
2011. A complex genomic rearrangement involving the endothelin 3 locus
causes dermal hyperpigmentation in the chicken. PLoS genetics 7,
e1002412.

Durkin, K., Coppieters, W., Drögemüller, C., Ahariz, N., Cambisano, N.,
Druet, T., Fasquelle, C., Haile, A., Horin, P., Huang, L., Kamatani, Y.,
Karim, L., Lathrop, M., Moser, S., Oldenbroek, K., Rieder, S., Sartelet,
A., Sölkner, J., Stålhammar, H., Zelenika, D., Zhang, Z., Leeb, T.,
Georges, M., Charlier, C., 2012. Serial translocation by means of
circular intermediates underlies colour sidedness in cattle. Nature 482,
81--84. https://doi.org/10.1038/nature10757

Ebert, P., Audano, P.A., Zhu, Q., Rodriguez-Martin, B., Porubsky, D.,
Bonder, M.J., Sulovari, A., Ebler, J., Zhou, W., Serra Mari, R., Yilmaz,
F., Zhao, X., Hsieh, P., Lee, J., Kumar, S., Lin, J., Rausch, T., Chen,
Y., Ren, J., Santamarina, M., Höps, W., Ashraf, H., Chuang, N.T., Yang,
X., Munson, K.M., Lewis, A.P., Fairley, S., Tallon, L.J., Clarke, W.E.,
Basile, A.O., Byrska-Bishop, M., Corvelo, A., Evani, U.S., Lu, T.-Y.,
Chaisson, M.J.P., Chen, J., Li, C., Brand, H., Wenger, A.M., Ghareghani,
M., Harvey, W.T., Raeder, B., Hasenfeld, P., Regier, A.A., Abel, H.J.,
Hall, I.M., Flicek, P., Stegle, O., Gerstein, M.B., Tubio, J.M.C., Mu,
Z., Li, Y.I., Shi, X., Hastie, A.R., Ye, K., Chong, Z., Sanders, A.D.,
Zody, M.C., Talkowski, M.E., Mills, R.E., Devine, S.E., Lee, C., Korbel,
J.O., Marschall, T., Eichler, E.E., 2021. Haplotype-resolved diverse
human genomes and integrated analysis of structural variation. Science
372, eabf7117. https://doi.org/10.1126/science.abf7117

Ebler, J., Ebert, P., Clarke, W.E., Rausch, T., Audano, P.A., Houwaart,
T., Mao, Y., Korbel, J.O., Eichler, E.E., Zody, M.C., Dilthey, A.T.,
Marschall, T., 2022. Pangenome-based genome inference allows efficient
and accurate genotyping across a wide spectrum of variant classes. Nat
Genet 54, 518--525. https://doi.org/10.1038/s41588-022-01043-w

Edriss, V., Fernando, R.L., Su, G., Lund, M.S., Guldbrandtsen, B., 2013.
The effect of using genealogy-based haplotypes for genomic prediction.
Genetics Selection Evolution 45, 5.
https://doi.org/10.1186/1297-9686-45-5

Erbe, M., Hayes, B.J., Matukumalli, L.K., Goswami, S., Bowman, P.J.,
Reich, C.M., Mason, B.A., Goddard, M.E., 2012. Improving accuracy of
genomic predictions within and between dairy cattle breeds with imputed
high-density single nucleotide polymorphism panels. Journal of Dairy
Science 95, 4114--4129. https://doi.org/10.3168/jds.2011-5019

Falconer, D.S., Mackay, T.F.C., 1996. Introduction to Quantitative
Genetics, Subsequent edition. ed. Benjamin-Cummings Pub Co, Harlow.

Fernando, R.L., Grossman, M., 1989. Marker assisted selection using best
linear unbiased prediction. Genetics Selection Evolution 21, 467--477.

Fragomeni, B.O., Lourenco, D.A., Masuda, Y., Legarra, A., Misztal, I.,
2017. Incorporation of causative quantitative trait nucleotides in
single-step GBLUP. Genetics Selection Evolution 49, 59.

Geibel, J., Praefke, N.P., Weigend, S., Simianer, H., Reimer, C., 2022.
Assessment of linkage disequilibrium patterns between structural
variants and single nucleotide polymorphisms in three commercial chicken
populations. BMC Genomics 23, 193.
https://doi.org/10.1186/s12864-022-08418-7

Geibel, J., Reimer, C., Pook, T., Weigend, S., Weigend, A., Simianer,
H., 2021. How imputation can mitigate SNP ascertainment Bias. BMC
Genomics 22, 340. https://doi.org/10.1186/s12864-021-07663-6

Georges, M., Charlier, C., Hayes, B., 2019. Harnessing genomic
information for livestock improvement. Nat Rev Genet 20, 135--156.

Giuffra, E., Tuggle, C.K., FAANG Consortium, 2019. Functional annotation
of animal genomes (FAANG): current achievements and roadmap. Annual
review of animal biosciences 7, 65--88.

Goddard, M., 2009. Genomic selection: prediction of accuracy and
maximisation of long term response. Genetica 136, 245--257.
https://doi.org/10.1007/s10709-008-9308-0

Goddard, M. e., Hayes, B. j., Meuwissen, T. h. e., 2011. Using the
genomic relationship matrix to predict the accuracy of genomic
selection. Journal of Animal Breeding and Genetics 128, 409--421.
https://doi.org/10.1111/j.1439-0388.2011.00964.x

Gunnarsson, U., Kerje, S., Bed'hom, B., Sahlqvist, A.-S., Ekwall, O.,
Tixier-Boichard, M., Kämpe, O., Andersson, L., 2011. The Dark brown
plumage color in chickens is caused by an 8.3-kb deletion upstream of
SOX10. Pigment Cell \& Melanoma Research 24, 268--274.
https://doi.org/10.1111/j.1755-148X.2011.00825.x

Habier, D., Fernando, R.L., Dekkers, J.C.M., 2007. The Impact of Genetic
Relationship Information on Genome-Assisted Breeding Values. Genetics
177, 2389--2397. https://doi.org/10.1534/genetics.107.081190

Habier, D., Fernando, R.L., Garrick, D.J., 2013. Genomic BLUP Decoded: A
Look into the Black Box of Genomic Prediction. Genetics 194, 597--607.
https://doi.org/10.1534/genetics.113.152207

Haley, C., Visscher, P., 1998. Strategies to utilize marker-quantitative
trait loci associations. Journal of dairy science 81, 85--97.

Halstead, M.M., Kern, C., Saelao, P., Wang, Y., Chanthavixay, G.,
Medrano, J.F., Van Eenennaam, A.L., Korf, I., Tuggle, C.K., Ernst, C.W.,
2020. A comparative analysis of chromatin accessibility in cattle, pig,
and mouse tissues. BMC Genomics 21, 698.

Hayes, B.J., Daetwyler, H.D., 2019. 1000 bull genomes project to map
simple and complex genetic traits in cattle: applications and outcomes.
Annual review of animal biosciences 7, 89--102.

Hickey, G., Heller, D., Monlong, J., Sibbesen, J.A., Sirén, J., Eizenga,
J., Dawson, E.T., Garrison, E., Novak, A.M., Paten, B., 2020. Genotyping
structural variants in pangenome graphs using the vg toolkit. Genome
Biol 21, 35. https://doi.org/10.1186/s13059-020-1941-7

Hickey, J. m., Kinghorn, B. p., Tier, B., Clark, S. a., van der Werf, J.
h. j., Gorjanc, G., 2013. Genomic evaluations using similarity between
haplotypes. Journal of Animal Breeding and Genetics 130, 259--269.
https://doi.org/10.1111/jbg.12020

Hickey, J.M., 2013. Sequencing millions of animals for genomic selection
2.0. Journal of Animal Breeding and Genetics 130, 331--332.
https://doi.org/10.1111/jbg.12054

Hickey, J.M., Chiurugwi, T., Mackay, I., Powell, W., 2017. Genomic
prediction unifies animal and plant breeding programs to form platforms
for biological discovery. Nat Genet 49, 1297--1303.
https://doi.org/10.1038/ng.3920

Ilska, J.J., 2015. Understanding genomic prediction in chickens.

Imsland, F., Feng, C., Boije, H., Bed'Hom, B., Fillon, V., Dorshorst,
B., Rubin, C.-J., Liu, R., Gao, Y., Gu, X., 2012. The Rose-comb mutation
in chickens constitutes a structural rearrangement causing both altered
comb morphology and defective sperm motility. PLoS genetics 8, e1002775.

Jang, S., Tsuruta, S., Leite, N.G., Misztal, I., Lourenco, D., 2023.
Dimensionality of genomic information and its impact on genome-wide
associations and variant selection for genomic prediction: a simulation
study. Genet Sel Evol 55, 49. https://doi.org/10.1186/s12711-023-00823-0

Jang, S., Tsuruta, S., Leite, N.G., Misztal, I., Lourenco, D., 2022.
Dimensionality of genomic information and its impact on GWA and variant
selection: a simulation study. https://doi.org/10.1101/2022.04.13.488175

Kadri, N.K., Sahana, G., Charlier, C., Iso-Touru, T., Guldbrandtsen, B.,
Karim, L., Nielsen, U.S., Panitz, F., Aamand, G.P., Schulman, N., 2014.
A 660-Kb deletion with antagonistic effects on fertility and milk
production segregates at high frequency in Nordic Red cattle: additional
evidence for the common occurrence of balancing selection in livestock.
PLoS Genet 10, e1004049.

Kelleher, J., Wong, Y., Wohns, A.W., Fadil, C., Albers, P.K., McVean,
G., 2019. Inferring whole-genome histories in large population datasets.
Nature genetics 51, 1330--1338.

Kern, C., Wang, Y., Xu, X., Pan, Z., Halstead, M., Chanthavixay, G.,
Saelao, P., Waters, S., Xiang, R., Chamberlain, A., Korf, I., Delany,
M.E., Cheng, H.H., Medrano, J.F., Van Eenennaam, A.L., Tuggle, C.K.,
Ernst, C., Flicek, P., Quon, G., Ross, P., Zhou, H., 2021. Functional
annotations of three domestic animal genomes provide vital resources for
comparative and agricultural research. Nat Commun 12, 1821.
https://doi.org/10.1038/s41467-021-22100-8

Knol, E.F., Nielsen, B., Knap, P.W., 2016. Genomic selection in
commercial pig breeding. Animal Frontiers 6, 15--22.

Lande, R., Thompson, R., 1990. Efficiency of marker-assisted selection
in the improvement of quantitative traits. Genetics 124, 743--756.

Legarra, A., Garcia-Baccino, C.A., Wientjes, Y.C.J., Vitezica, Z.G.,
2021. The correlation of substitution effects across populations and
generations in the presence of nonadditive functional gene action.
Genetics 219, iyab138. https://doi.org/10.1093/genetics/iyab138

Leonard, A.S., Crysnanto, D., Fang, Z.-H., Heaton, M.P., Vander Ley,
B.L., Herrera, C., Bollwein, H., Bickhart, D.M., Kuhn, K.L., Smith,
T.P.L., Rosen, B.D., Pausch, H., 2022. Structural variant-based
pangenome construction has low sensitivity to variability of
haplotype-resolved bovine assemblies. Nat Commun 13, 3012.
https://doi.org/10.1038/s41467-022-30680-2

Littlejohn, M., Lopdell, T., Trevarton, A., Moody, J., Tiplady, K.,
Burborough, K., Prowse-Wilkins, C., Chamberlain, A., Goddard, M., Snell,
R., 2022. A massively parallel reporter assay to screen bovine
regulatory variants. Presented at the World Congress of Genetics Applied
to Livestock Production 2022.

Liu, L., Sanderford, M.D., Patel, R., Chandrashekar, P., Gibson, G.,
Kumar, S., 2019. Biological relevance of computationally predicted
pathogenicity of noncoding variants. Nat Commun 10, 330.
https://doi.org/10.1038/s41467-018-08270-y

Liu, S., Gao, Y., Canela-Xandri, O., Wang, S., Yu, Y., Cai, W., Li, B.,
Xiang, R., Chamberlain, A.J., Pairo-Castineira, E., D'Mellow, K.,
Rawlik, K., Xia, C., Yao, Y., Navarro, P., Rocha, D., Li, X., Yan, Z.,
Li, C., Rosen, B.D., Van Tassell, C.P., Vanraden, P.M., Zhang, S., Ma,
L., Cole, J.B., Liu, G.E., Tenesa, A., Fang, L., 2022. A multi-tissue
atlas of regulatory variants in cattle. Nat Genet 54, 1438--1447.
https://doi.org/10.1038/s41588-022-01153-5

Lowe, J.W., Bruce, A., 2019. Genetics without genes? The centrality of
genetic markers in livestock genetics and genomics. History and
philosophy of the life sciences 41, 50.

MacLeod, I.M., Hayes, B.J., Goddard, M.E., 2014. The effects of
demography and long-term selection on the accuracy of genomic prediction
with sequence data. Genetics 198, 1671--1684.

Meuwissen, T., Goddard, M., 2010. Accurate Prediction of Genetic Values
for Complex Traits by Whole-Genome Resequencing. Genetics 185, 623--631.
https://doi.org/10.1534/genetics.110.116590

Meuwissen, T., van den Berg, I., Goddard, M., 2021. On the use of
whole-genome sequence data for across-breed genomic prediction and
fine-scale mapping of QTL. Genetics Selection Evolution 53, 19.
https://doi.org/10.1186/s12711-021-00607-4

Meuwissen, T.H.E., Hayes, B., Goddard, M., 2001. Prediction of total
genetic value using genome-wide dense marker maps. Genetics 157,
1819--1829.

Mishra, N.A., Drögemüller, C., Jagannathan, V., Keller, I., Wüthrich,
D., Bruggmann, R., Beck, J., Schütz, E., Brenig, B., Demmel, S., Moser,
S., Signer-Hasler, H., Pieńkowska-Schelling, A., Schelling, C., Sande,
M., Rongen, R., Rieder, S., Kelsh, R.N., Mercader, N., Leeb, T., 2017. A
structural variant in the 5'-flanking region of the TWIST2 gene affects
melanocyte development in belted cattle. PLOS ONE 12, e0180170.
https://doi.org/10.1371/journal.pone.0180170

Misztal, I., Lourenco, D., Legarra, A., 2020. Current status of genomic
evaluation. Journal of Animal Science 98, skaa101.
https://doi.org/10.1093/jas/skaa101

Misztal, I., Steyn, Y., Lourenco, D. a. L., 2022. Genomic evaluation
with multibreed and crossbred data *. JDS Communications 3, 156--159.
https://doi.org/10.3168/jdsc.2021-0177

Moghaddar, N., Khansefid, M., van der Werf, J.H.J., Bolormaa, S.,
Duijvesteijn, N., Clark, S.A., Swan, A.A., Daetwyler, H.D., MacLeod,
I.M., 2019. Genomic prediction based on selected variants from imputed
whole-genome sequence data in Australian sheep populations. Genetics
Selection Evolution 51, 72. https://doi.org/10.1186/s12711-019-0514-2

Mullen, M.P., McClure, M.C., Kearney, J.F., Waters, S.M., Weld, R.,
Flynn, P., Creevey, C.J., Cromie, A.R., Berry, D.P., 2013. Development
of a custom SNP chip for dairy and beef cattle breeding, parentage and
research. Interbull Bulletin.

Nejati-Javaremi, A., Smith, C., Gibson, J., 1997. Effect of total
allelic relationship on accuracy of evaluation and response to
selection. Journal of animal science 75, 1738--1745.

Nguyen, T.V., Vander Jagt, C.J., Wang, J., Daetwyler, H.D., Xiang, R.,
Goddard, M.E., Nguyen, L.T., Ross, E.M., Hayes, B.J., Chamberlain, A.J.,
MacLeod, I.M., 2023. In it for the long run: perspectives on exploiting
long-read sequencing in livestock for population scale studies of
structural variants. Genetics Selection Evolution 55, 9.
https://doi.org/10.1186/s12711-023-00783-5

Oppong, R.F., Boutin, T., Campbell, A., McIntosh, A.M., Porteous, D.,
Hayward, C., Haley, C.S., Navarro, P., Knott, S., 2022. SNP and
Haplotype Regional Heritability Mapping (SNHap-RHM): Joint Mapping of
Common and Rare Variation Affecting Complex Traits. Frontiers in
Genetics 12.

Pérez-Enciso, M., Rincón, J.C., Legarra, A., 2015. Sequence- vs.
chip-assisted genomic selection: accurate biological information is
advised. Genetics Selection Evolution 47, 43.
https://doi.org/10.1186/s12711-015-0117-5

Pocrnic, I., Lourenco, D.A., Masuda, Y., Legarra, A., Misztal, I.,
2016a. The dimensionality of genomic information and its effect on
genomic prediction. Genetics 203, 573--581.

Pocrnic, I., Lourenco, D.A.L., Masuda, Y., Misztal, I., 2019. Accuracy
of genomic BLUP when considering a genomic relationship matrix based on
the number of the largest eigenvalues: a simulation study. Genetics
Selection Evolution 51, 75. https://doi.org/10.1186/s12711-019-0516-0

Pocrnic, I., Lourenco, D.A.L., Masuda, Y., Misztal, I., 2016b.
Dimensionality of genomic information and performance of the Algorithm
for Proven and Young for different livestock species. Genetics Selection
Evolution 48, 82. https://doi.org/10.1186/s12711-016-0261-6

Pook, T., Freudenthal, J., Korte, A., Simianer, H., 2020. Using Local
Convolutional Neural Networks for Genomic Prediction. Frontiers in
Genetics 11.

Pook, T., Schlather, M., de los Campos, G., Mayer, M., Schoen, C.C.,
Simianer, H., 2019. HaploBlocker: Creation of Subgroup-Specific
Haplotype Blocks and Libraries. Genetics 212, 1045--1061.
https://doi.org/10.1534/genetics.119.302283

Prowse-Wilkins, C.P., Lopdell, T.J., Xiang, R., Vander Jagt, C.J.,
Littlejohn, M.D., Chamberlain, A.J., Goddard, M.E., 2022. Genetic
variation in histone modifications and gene expression identifies
regulatory variants in the mammary gland of cattle. BMC Genomics 23,
815. https://doi.org/10.1186/s12864-022-09002-9

Ragoussis, J., 2009. Genotyping Technologies for Genetic Research.
Annual Review of Genomics and Human Genetics 10, 117--133.
https://doi.org/10.1146/annurev-genom-082908-150116

Raymond, B., Bouwman, A.C., Schrooten, C., Houwing-Duistermaat, J.,
Veerkamp, R.F., 2018a. Utility of whole-genome sequence data for
across-breed genomic prediction. Genet Sel Evol 50, 27.
https://doi.org/10.1186/s12711-018-0396-8

Raymond, B., Bouwman, A.C., Wientjes, Y.C.J., Schrooten, C.,
Houwing-Duistermaat, J., Veerkamp, R.F., 2018b. Genomic prediction for
numerically small breeds, using models with pre-selected and
differentially weighted markers. Genetics Selection Evolution 50, 49.
https://doi.org/10.1186/s12711-018-0419-5

Ros-Freixedes, R., Battagin, M., Johnsson, M., Gorjanc, G., Mileham,
A.J., Rounsley, S.D., Hickey, J.M., 2018. Impact of index hopping and
bias towards the reference allele on accuracy of genotype calls from
low-coverage sequencing. Genetics Selection Evolution 50, 64.
https://doi.org/10.1186/s12711-018-0436-4

Ros-Freixedes, R., Johnsson, M., Whalen, A., Chen, C.-Y., Valente, B.D.,
Herring, W.O., Gorjanc, G., Hickey, J.M., 2022a. Genomic prediction with
whole-genome sequence data in intensely selected pig lines. Genetics
Selection Evolution 54, 65. https://doi.org/10.1186/s12711-022-00756-0

Ros-Freixedes, R., Valente, B.D., Chen, C.-Y., Herring, W.O., Gorjanc,
G., Hickey, J.M., Johnsson, M., 2022b. Rare and population-specific
functional variation across pig lines. Genetics Selection Evolution 54,
39. https://doi.org/10.1186/s12711-022-00732-8

Ros-Freixedes, R., Whalen, A., Chen, C.-Y., Gorjanc, G., Herring, W.O.,
Mileham, A.J., Hickey, J.M., 2020. Accuracy of whole-genome sequence
imputation using hybrid peeling in large pedigreed livestock
populations. Genetics Selection Evolution 52, 17.
https://doi.org/10.1186/s12711-020-00536-8

Rubin, C.-J., Megens, H.-J., Barrio, A.M., Maqbool, K., Sayyab, S.,
Schwochow, D., Wang, C., Carlborg, Ö., Jern, P., Jørgensen, C.B., 2012.
Strong signatures of selection in the domestic pig genome. Proc Natl
Acad Sci USA 109, 19529--19536.

Salavati, M., Woolley, S.A., Cortés Araya, Y., Halstead, M.M.,
Stenhouse, C., Johnsson, M., Ashworth, C.J., Archibald, A.L., Donadeu,
F.X., Hassan, M.A., Clark, E.L., 2022. Profiling of open chromatin in
developing pig (Sus scrofa) muscle to identify regulatory regions. G3
Genes\textbar Genomes\textbar Genetics 12, jkab424.
https://doi.org/10.1093/g3journal/jkab424

Schütz, E., Scharfenstein, M., Brenig, B., 2008. Implication of Complex
Vertebral Malformation and Bovine Leukocyte Adhesion Deficiency
DNA-Based Testing on Disease Frequency in the Holstein Population.
Journal of Dairy Science 91, 4854--4859.
https://doi.org/10.3168/jds.2008-1154

Selle, M.L., Steinsland, I., Lindgren, F., Brajkovic, V., Cubric-Curik,
V., Gorjanc, G., 2021. Hierarchical Modelling of Haplotype Effects on a
Phylogeny. Frontiers in Genetics 11.

Smith, C., 1967. Improvement of metric traits through specific genetic
loci. Animal Science 9, 349--358.
https://doi.org/10.1017/S0003356100038642

Snelling, W.M., Hoff, J.L., Li, J.H., Kuehn, L.A., Keel, B.N.,
Lindholm-Perry, A.K., Pickrell, J.K., 2020. Assessment of imputation
from low-pass sequencing to predict merit of beef steers. Genes 11,
1312.

Soller, M., 1978. The use of loci associated with quantitative effects
in dairy cattle improvement. Animal Science 27, 133--139.
https://doi.org/10.1017/S0003356100035960

Stam, P., 1980. The distribution of the fraction of the genome identical
by descent in finite random mating populations. Genetics Research 35,
131--155. https://doi.org/10.1017/S0016672300014002

Sudmant, P.H., Rausch, T., Gardner, E.J., Handsaker, R.E., Abyzov, A.,
Huddleston, J., Zhang, Y., Ye, K., Jun, G., Hsi-Yang Fritz, M., Konkel,
M.K., Malhotra, A., Stütz, A.M., Shi, X., Paolo Casale, F., Chen, J.,
Hormozdiari, F., Dayama, G., Chen, K., Malig, M., Chaisson, M.J.P.,
Walter, K., Meiers, S., Kashin, S., Garrison, E., Auton, A., Lam,
H.Y.K., Jasmine Mu, X., Alkan, C., Antaki, D., Bae, T., Cerveira, E.,
Chines, P., Chong, Z., Clarke, L., Dal, E., Ding, L., Emery, S., Fan,
X., Gujral, M., Kahveci, F., Kidd, J.M., Kong, Y., Lameijer, E.-W.,
McCarthy, S., Flicek, P., Gibbs, R.A., Marth, G., Mason, C.E., Menelaou,
A., Muzny, D.M., Nelson, B.J., Noor, A., Parrish, N.F., Pendleton, M.,
Quitadamo, A., Raeder, B., Schadt, E.E., Romanovitch, M., Schlattl, A.,
Sebra, R., Shabalin, A.A., Untergasser, A., Walker, J.A., Wang, M., Yu,
F., Zhang, C., Zhang, J., Zheng-Bradley, X., Zhou, W., Zichner, T.,
Sebat, J., Batzer, M.A., McCarroll, S.A., Mills, R.E., Gerstein, M.B.,
Bashir, A., Stegle, O., Devine, S.E., Lee, C., Eichler, E.E., Korbel,
J.O., 2015. An integrated map of structural variation in 2,504 human
genomes. Nature 526, 75--81. https://doi.org/10.1038/nature15394

Sved, J., 1971. Linkage disequilibrium and homozygosity of chromosome
segments in finite populations. Theoretical population biology 2,
125--141.

Talenti, A., Powell, J., Hemmink, J.D., Cook, E. a. J., Wragg, D.,
Jayaraman, S., Paxton, E., Ezeasor, C., Obishakin, E.T., Agusi, E.R.,
Tijjani, A., Amanyire, W., Muhanguzi, D., Marshall, K., Fisch, A.,
Ferreira, B.R., Qasim, A., Chaudhry, U., Wiener, P., Toye, P., Morrison,
L.J., Connelley, T., Prendergast, J.G.D., 2022. A cattle graph genome
incorporating global breed diversity. Nat Commun 13, 910.
https://doi.org/10.1038/s41467-022-28605-0

The FarmGTEx-PigGTEx Consortium, Y., Yin, H., Bai, Z., Liu, S., Zeng,
H., Bai, L., Cai, Z., Zhao, B., Li, X., Xu, Z., Lin, Q., Pan, Z., Yang,
W., Yu, X., Guan, D., Hou, Y., Keel, B.N., Rohrer, G.A., Lindholm-Perry,
A.K., Oliver, W.T., Ballester, M., Crespo-Piazuelo, D., Quintanilla, R.,
Canela-Xandri, O., Rawlik, K., Xia, C., Yao, Y., Zhao, Q., Yao, W.,
Yang, L., Li, H., Zhang, H., Liao, W., Chen, T., Karlskov-Mortensen, P.,
Fredholm, M., Amills, M., Clop, A., Giuffra, E., Wu, J., Cai, X., Diao,
S., Pan, X., Wei, C., Li, Jinghui, Cheng, H., Wang, S., Su, G., Sahana,
G., Lund, M.S., Dekkers, J.C.M., Kramer, L., Tuggle, C.K., Corbett, R.,
Groenen, M.A.M., Madsen, O., Gòdia, M., Rocha, D., Charles, M., Li, C.,
Pausch, H., Hu, X., Frantz, L., Luo, Y., Lin, L., Zhou, Z., Zhang, Z.,
Chen, Z., Cui, L., Xiang, R., Shen, X., Li, P., Huang, R., Tang, G., Li,
M., Zhao, Y., Yi, G., Tang, Z., Jiang, J., Zhao, F., Yuan, X., Liu, X.,
Chen, Y., Xu, X., Zhao, S., Zhao, P., Haley, C., Zhou, H., Wang, Q.,
Pan, Y., Ding, X., Ma, L., Li, Jiaqi, Navarro, P., Zhang, Q., Li, B.,
Tenesa, A., Li, K., Liu, G.E., Zhang, Z., Fang, L., 2022. A compendium
of genetic regulatory effects across pig tissues.
https://doi.org/10.1101/2022.11.11.516073

Tian, X., Li, R., Fu, W., Li, Y., Wang, X., Li, M., Du, D., Tang, Q.,
Cai, Y., Long, Y., 2019. Building a sequence map of the pig pan-genome
from multiple de novo assemblies and Hi-C data. Sci China Life Sci
750--63.

van Binsbergen, R., Calus, M.P.L., Bink, M.C.A.M., van Eeuwijk, F.A.,
Schrooten, C., Veerkamp, R.F., 2015. Genomic prediction using imputed
whole-genome sequence data in Holstein Friesian cattle. Genetics
Selection Evolution 47, 71. https://doi.org/10.1186/s12711-015-0149-x

van den Berg, I., Bowman, P.J., MacLeod, I.M., Hayes, B.J., Wang, T.,
Bolormaa, S., Goddard, M.E., 2017. Multi-breed genomic prediction using
Bayes R with sequence data and dropping variants with a small effect.
Genet Sel Evol 49, 70. https://doi.org/10.1186/s12711-017-0347-9

Van der Auwera, G.A., Carneiro, M.O., Hartl, C., Poplin, R., del Angel,
G., Levy-Moonshine, A., Jordan, T., Shakir, K., Roazen, D., Thibault,
J., Banks, E., Garimella, K.V., Altshuler, D., Gabriel, S., DePristo,
M.A., 2013. From FastQ Data to High-Confidence Variant Calls: The Genome
Analysis Toolkit Best Practices Pipeline. Current Protocols in
Bioinformatics 43, 11.10.1-11.10.33.
https://doi.org/10.1002/0471250953.bi1110s43

VanRaden, P.M., 2008. Efficient Methods to Compute Genomic Predictions.
Journal of Dairy Science 91, 4414--4423.
https://doi.org/10.3168/jds.2007-0980

VanRaden, P.M., Tooker, M.E., O'connell, J.R., Cole, J.B., Bickhart,
D.M., 2017. Selecting sequence variants to improve genomic predictions
for dairy cattle. Genetics Selection Evolution 49, 1--12.

Veerkamp, R.F., Bouwman, A.C., Schrooten, C., Calus, M.P.L., 2016.
Genomic prediction using preselected DNA variants from a GWAS with
whole-genome sequence data in Holstein--Friesian cattle. Genet Sel Evol
48, 95. https://doi.org/10.1186/s12711-016-0274-1

Wang, M., Hancock, T.P., MacLeod, I.M., Pryce, J.E., Cocks, B.G., Hayes,
B.J., 2017. Putative enhancer sites in the bovine genome are enriched
with variants affecting complex traits. Genetics Selection Evolution 49,
56. https://doi.org/10.1186/s12711-017-0331-4

Wang, Z., Qu, L., Yao, J., Yang, X., Li, G., Zhang, Y., Li, J., Wang,
X., Bai, J., Xu, G., 2013. An EAV-HP insertion in 5' flanking region of
SLCO1B3 causes blue eggshell in the chicken. Plos genetics 9, e1003183.

Whalen, A., Ros-Freixedes, R., Wilson, D.L., Gorjanc, G., Hickey, J.M.,
2018. Hybrid peeling for fast and accurate calling, phasing, and
imputation with sequence data of any coverage in pedigrees. Genetics
Selection Evolution 50, 67.

Wiedemar, N., Tetens, J., Jagannathan, V., Menoud, A., Neuenschwander,
S., Bruggmann, R., Thaller, G., Drögemüller, C., 2014. Independent
Polled Mutations Leading to Complex Gene Expression Differences in
Cattle. PLOS ONE 9, e93435. https://doi.org/10.1371/journal.pone.0093435

Wientjes, Y.C., Calus, M.P., Goddard, M.E., Hayes, B.J., 2015. Impact of
QTL properties on the accuracy of multi-breed genomic prediction.
Genetics Selection Evolution 47, 42.
https://doi.org/10.1186/s12711-015-0124-6

Wientjes, Y.C.J., Veerkamp, R.F., Calus, M.P.L., 2013. The Effect of
Linkage Disequilibrium and Family Relationships on the Reliability of
Genomic Prediction. Genetics 193, 621--631.
https://doi.org/10.1534/genetics.112.146290

Wiggans, G.R., Cole, J.B., Hubbard, S.M., Sonstegard, T.S., 2017.
Genomic Selection in Dairy Cattle: The USDA Experience. Annual Review of
Animal Biosciences 5, 309--327.
https://doi.org/10.1146/annurev-animal-021815-111422

Wolc, A., Kranis, A., Arango, J., Settar, P., Fulton, J., O'Sullivan,
N., Avendano, A., Watson, K., Hickey, J., De los Campos, G., 2016.
Implementation of genomic selection in the poultry industry. Animal
Frontiers 6, 23--31.

Wright, D., Boije, H., Meadows, J.R.S., Bed'Hom, B., Gourichon, D.,
Vieaud, A., Tixier-Boichard, M., Rubin, C.-J., Imsland, F., Hallböök,
F., 2009. Copy number variation in intron 1 of SOX5 causes the Pea-comb
phenotype in chickens. PLoS genetics 5, e1000512.

Xiang, R., Fang, L., Liu, S., Liu, G.E., Tenesa, A., Gao, Y.,
Consortium, C., Mason, B.A., Chamberlain, A.J., Goddard, M.E., 2022a.
Genetic score omics regression and multi-trait meta-analysis detect
widespread cis-regulatory effects shaping bovine complex traits.
https://doi.org/10.1101/2022.07.13.499886

Xiang, R., Fang, L., Liu, S., Macleod, I.M., Liu, Z., Breen, E.J., Gao,
Y., Liu, G.E., Tenesa, A., Consortium, C., Mason, B.A., Chamberlain,
A.J., Wray, N.R., Goddard, M.E., 2022b. Gene expression and RNA splicing
explain large proportions of the heritability for complex traits in
cattle. https://doi.org/10.1101/2022.05.30.494093

Xiang, R., MacLeod, I.M., Daetwyler, H.D., de Jong, G., O'Connor, E.,
Schrooten, C., Chamberlain, A.J., Goddard, M.E., 2021. Genome-wide
fine-mapping identifies pleiotropic and functional variants that predict
many traits across global cattle populations. Nature communications 12,
1--13.

Xiang, R., van den Berg, I., MacLeod, I.M., Hayes, B.J., Prowse-Wilkins,
C.P., Wang, M., Bolormaa, S., Liu, Z., Rochfort, S.J., Reich, C.M.,
Mason, B.A., Vander Jagt, C.J., Daetwyler, H.D., Lund, M.S.,
Chamberlain, A.J., Goddard, M.E., 2019. Quantifying the contribution of
sequence variants with regulatory and evolutionary significance to 34
bovine complex traits. Proc Natl Acad Sci USA 116, 19398--408.
https://doi.org/10.1073/pnas.1904159116

Xu, L., Cole, J.B., Bickhart, D.M., Hou, Y., Song, J., VanRaden, P.M.,
Sonstegard, T.S., Van Tassell, C.P., Liu, G.E., 2014. Genome wide CNV
analysis reveals additional variants associated with milk production
traits in Holsteins. BMC Genomics 15, 683.
https://doi.org/10.1186/1471-2164-15-683

Yan, S.M., Sherman, R.M., Taylor, D.J., Nair, D.R., Bortvin, A.N.,
Schatz, M.C., McCoy, R.C., 2021. Local adaptation and archaic
introgression shape global diversity at human structural variant loci.
eLife 10, e67615. https://doi.org/10.7554/eLife.67615

Yengo, L., Vedantam, S., Marouli, E., Sidorenko, J., Bartell, E.,
Sakaue, S., Graff, M., Eliasen, A.U., Jiang, Y., Raghavan, S., Miao, J.,
Arias, J.D., Graham, S.E., Mukamel, R.E., Spracklen, C.N., Yin, X.,
Chen, S.-H., Ferreira, T., Highland, H.H., Ji, Y., Karaderi, T., Lin,
K., Lüll, K., Malden, D.E., Medina-Gomez, C., Machado, M., Moore, A.,
Rüeger, S., Sim, X., Vrieze, S., Ahluwalia, T.S., Akiyama, M., Allison,
M.A., Alvarez, M., Andersen, M.K., Ani, A., Appadurai, V., Arbeeva, L.,
Bhaskar, S., Bielak, L.F., Bollepalli, S., Bonnycastle, L.L.,
Bork-Jensen, J., Bradfield, J.P., Bradford, Y., Braund, P.S., Brody,
J.A., Burgdorf, K.S., Cade, B.E., Cai, H., Cai, Q., Campbell, A.,
Cañadas-Garre, M., Catamo, E., Chai, J.-F., Chai, X., Chang, L.-C.,
Chang, Y.-C., Chen, C.-H., Chesi, A., Choi, S.H., Chung, R.-H., Cocca,
M., Concas, M.P., Couture, C., Cuellar-Partida, G., Danning, R., Daw,
E.W., Degenhard, F., Delgado, G.E., Delitala, A., Demirkan, A., Deng,
X., Devineni, P., Dietl, A., Dimitriou, M., Dimitrov, L., Dorajoo, R.,
Ekici, A.B., Engmann, J.E., Fairhurst-Hunter, Z., Farmaki, A.-E., Faul,
J.D., Fernandez-Lopez, J.-C., Forer, L., Francescatto, M., Freitag-Wolf,
S., Fuchsberger, C., Galesloot, T.E., Gao, Y., Gao, Z., Geller, F.,
Giannakopoulou, O., Giulianini, F., Gjesing, A.P., Goel, A., Gordon,
S.D., Gorski, M., Grove, J., Guo, X., Gustafsson, S., Haessler, J.,
Hansen, T.F., Havulinna, A.S., Haworth, S.J., He, J., Heard-Costa, N.,
Hebbar, P., Hindy, G., Ho, Y.-L.A., Hofer, E., Holliday, E., Horn, K.,
Hornsby, W.E., Hottenga, J.-J., Huang, H., Huang, J., Huerta-Chagoya,
A., Huffman, J.E., Hung, Y.-J., Huo, S., Hwang, M.Y., Iha, H., Ikeda,
D.D., Isono, M., Jackson, A.U., Jäger, S., Jansen, I.E., Johansson, I.,
Jonas, J.B., Jonsson, A., Jørgensen, T., Kalafati, I.-P., Kanai, M.,
Kanoni, S., Kårhus, L.L., Kasturiratne, A., Katsuya, T., Kawaguchi, T.,
Kember, R.L., Kentistou, K.A., Kim, H.-N., Kim, Y.J., Kleber, M.E.,
Knol, M.J., Kurbasic, A., Lauzon, M., Le, P., Lea, R., Lee, J.-Y.,
Leonard, H.L., Li, S.A., Li, Xiaohui, Li, Xiaoyin, Liang, J., Lin, H.,
Lin, S.-Y., Liu, Jun, Liu, X., Lo, K.S., Long, J., Lores-Motta, L.,
Luan, J., Lyssenko, V., Lyytikäinen, L.-P., Mahajan, A., Mamakou, V.,
Mangino, M., Manichaikul, A., Marten, J., Mattheisen, M., Mavarani, L.,
McDaid, A.F., Meidtner, K., Melendez, T.L., Mercader, J.M., Milaneschi,
Y., Miller, J.E., Millwood, I.Y., Mishra, P.P., Mitchell, R.E.,
Møllehave, L.T., Morgan, A., Mucha, S., Munz, M., Nakatochi, M., Nelson,
C.P., Nethander, M., Nho, C.W., Nielsen, A.A., Nolte, I.M., Nongmaithem,
S.S., Noordam, R., Ntalla, I., Nutile, T., Pandit, A., Christofidou, P.,
Pärna, K., Pauper, M., Petersen, E.R.B., Petersen, L.V., Pitkänen, N.,
Polašek, O., Poveda, A., Preuss, M.H., Pyarajan, S., Raffield, L.M.,
Rakugi, H., Ramirez, J., Rasheed, A., Raven, D., Rayner, N.W., Riveros,
C., Rohde, R., Ruggiero, D., Ruotsalainen, S.E., Ryan, K.A.,
Sabater-Lleal, M., Saxena, R., Scholz, M., Sendamarai, A., Shen, B.,
Shi, J., Shin, J.H., Sidore, C., Sitlani, C.M., Slieker, R.C., Smit,
R.A.J., Smith, A.V., Smith, J.A., Smyth, L.J., Southam, L.,
Steinthorsdottir, V., Sun, L., Takeuchi, F., Tallapragada, D.S.P.,
Taylor, K.D., Tayo, B.O., Tcheandjieu, C., Terzikhan, N., Tesolin, P.,
Teumer, A., Theusch, E., Thompson, D.J., Thorleifsson, G., Timmers,
P.R.H.J., Trompet, S., Turman, C., Vaccargiu, S., van der Laan, S.W.,
van der Most, P.J., van Klinken, J.B., van Setten, J., Verma, S.S.,
Verweij, N., Veturi, Y., Wang, C.A., Wang, C., Wang, L., Wang, Z.,
Warren, H.R., Bin Wei, W., Wickremasinghe, A.R., Wielscher, M., Wiggins,
K.L., Winsvold, B.S., Wong, A., Wu, Y., Wuttke, M., Xia, R., Xie, T.,
Yamamoto, K., Yang, Jingyun, Yao, J., Young, H., Yousri, N.A., Yu, L.,
Zeng, L., Zhang, W., Zhang, X., Zhao, J.-H., Zhao, W., Zhou, W.,
Zimmermann, M.E., Zoledziewska, M., Adair, L.S., Adams, H.H.H.,
Aguilar-Salinas, C.A., Al-Mulla, F., Arnett, D.K., Asselbergs, F.W.,
Åsvold, B.O., Attia, J., Banas, B., Bandinelli, S., Bennett, D.A.,
Bergler, T., Bharadwaj, D., Biino, G., Bisgaard, H., Boerwinkle, E.,
Böger, C.A., Bønnelykke, K., Boomsma, D.I., Børglum, A.D., Borja, J.B.,
Bouchard, C., Bowden, D.W., Brandslund, I., Brumpton, B., Buring, J.E.,
Caulfield, M.J., Chambers, J.C., Chandak, G.R., Chanock, S.J.,
Chaturvedi, N., Chen, Y.-D.I., Chen, Z., Cheng, C.-Y., Christophersen,
I.E., Ciullo, M., Cole, J.W., Collins, F.S., Cooper, R.S., Cruz, M.,
Cucca, F., Cupples, L.A., Cutler, M.J., Damrauer, S.M., Dantoft, T.M.,
de Borst, G.J., de Groot, L.C.P.G.M., De Jager, P.L., de Kleijn, D.P.V.,
Janaka de Silva, H., Dedoussis, G.V., den Hollander, A.I., Du, S.,
Easton, D.F., Elders, P.J.M., Eliassen, A.H., Ellinor, P.T., Elmståhl,
S., Erdmann, J., Evans, M.K., Fatkin, D., Feenstra, B., Feitosa, M.F.,
Ferrucci, L., Ford, I., Fornage, M., Franke, A., Franks, P.W., Freedman,
B.I., Gasparini, P., Gieger, C., Girotto, G., Goddard, M.E., Golightly,
Y.M., Gonzalez-Villalpando, C., Gordon-Larsen, P., Grallert, H., Grant,
S.F.A., Grarup, N., Griffiths, L., Gudnason, V., Haiman, C., Hakonarson,
H., Hansen, T., Hartman, C.A., Hattersley, A.T., Hayward, C., Heckbert,
S.R., Heng, C.-K., Hengstenberg, C., Hewitt, A.W., Hishigaki, H., Hoyng,
C.B., Huang, P.L., Huang, W., Hunt, S.C., Hveem, K., Hyppönen, E.,
Iacono, W.G., Ichihara, S., Ikram, M.A., Isasi, C.R., Jackson, R.D.,
Jarvelin, M.-R., Jin, Z.-B., Jöckel, K.-H., Joshi, P.K., Jousilahti, P.,
Jukema, J.W., Kähönen, M., Kamatani, Y., Kang, K.D., Kaprio, J., Kardia,
S.L.R., Karpe, F., Kato, N., Kee, F., Kessler, T., Khera, A.V., Khor,
C.C., Kiemeney, L.A.L.M., Kim, B.-J., Kim, E.K., Kim, H.-L., Kirchhof,
P., Kivimaki, M., Koh, W.-P., Koistinen, H.A., Kolovou, G.D., Kooner,
J.S., Kooperberg, C., Köttgen, A., Kovacs, P., Kraaijeveld, A., Kraft,
P., Krauss, R.M., Kumari, M., Kutalik, Z., Laakso, M., Lange, L.A.,
Langenberg, C., Launer, L.J., Le Marchand, L., Lee, H., Lee, N.R.,
Lehtimäki, T., Li, H., Li, L., Lieb, W., Lin, X., Lind, L., Linneberg,
A., Liu, C.-T., Liu, Jianjun, Loeffler, M., London, B., Lubitz, S.A.,
Lye, S.J., Mackey, D.A., Mägi, R., Magnusson, P.K.E., Marcus, G.M.,
Vidal, P.M., Martin, N.G., März, W., Matsuda, F., McGarrah, R.W., McGue,
M., McKnight, A.J., Medland, S.E., Mellström, D., Metspalu, A.,
Mitchell, B.D., Mitchell, P., Mook-Kanamori, D.O., Morris, A.D., Mucci,
L.A., Munroe, P.B., Nalls, M.A., Nazarian, S., Nelson, A.E., Neville,
M.J., Newton-Cheh, C., Nielsen, C.S., Nöthen, M.M., Ohlsson, C.,
Oldehinkel, A.J., Orozco, L., Pahkala, K., Pajukanta, P., Palmer,
C.N.A., Parra, E.J., Pattaro, C., Pedersen, O., Pennell, C.E., Penninx,
B.W.J.H., Perusse, L., Peters, A., Peyser, P.A., Porteous, D.J.,
Posthuma, D., Power, C., Pramstaller, P.P., Province, M.A., Qi, Q., Qu,
J., Rader, D.J., Raitakari, O.T., Ralhan, S., Rallidis, L.S., Rao, D.C.,
Redline, S., Reilly, D.F., Reiner, A.P., Rhee, S.Y., Ridker, P.M.,
Rienstra, M., Ripatti, S., Ritchie, M.D., Roden, D.M., Rosendaal, F.R.,
Rotter, J.I., Rudan, I., Rutters, F., Sabanayagam, C., Saleheen, D.,
Salomaa, V., Samani, N.J., Sanghera, D.K., Sattar, N., Schmidt, B.,
Schmidt, H., Schmidt, R., Schulze, M.B., Schunkert, H., Scott, L.J.,
Scott, R.J., Sever, P., Shiroma, E.J., Shoemaker, M.B., Shu, X.-O.,
Simonsick, E.M., Sims, M., Singh, J.R., Singleton, A.B., Sinner, M.F.,
Smith, J.G., Snieder, H., Spector, T.D., Stampfer, M.J., Stark, K.J.,
Strachan, D.P., `t Hart, L.M., Tabara, Y., Tang, H., Tardif, J.-C.,
Thanaraj, T.A., Timpson, N.J., Tönjes, A., Tremblay, A., Tuomi, T.,
Tuomilehto, J., Tusié-Luna, M.-T., Uitterlinden, A.G., van Dam, R.M.,
van der Harst, P., Van der Velde, N., van Duijn, C.M., van Schoor, N.M.,
Vitart, V., Völker, U., Vollenweider, P., Völzke, H., Wacher-Rodarte,
N.H., Walker, M., Wang, Y.X., Wareham, N.J., Watanabe, R.M., Watkins,
H., Weir, D.R., Werge, T.M., Widen, E., Wilkens, L.R., Willemsen, G.,
Willett, W.C., Wilson, J.F., Wong, T.-Y., Woo, J.-T., Wright, A.F., Wu,
J.-Y., Xu, H., Yajnik, C.S., Yokota, M., Yuan, J.-M., Zeggini, E.,
Zemel, B.S., Zheng, W., Zhu, X., Zmuda, J.M., Zonderman, A.B., Zwart,
J.-A., Chasman, D.I., Cho, Y.S., Heid, I.M., McCarthy, M.I., Ng, M.C.Y.,
O'Donnell, C.J., Rivadeneira, F., Thorsteinsdottir, U., Sun, Y.V., Tai,
E.S., Boehnke, M., Deloukas, P., Justice, A.E., Lindgren, C.M., Loos,
R.J.F., Mohlke, K.L., North, K.E., Stefansson, K., Walters, R.G.,
Winkler, T.W., Young, K.L., Loh, P.-R., Yang, Jian, Esko, T., Assimes,
T.L., Auton, A., Abecasis, G.R., Willer, C.J., Locke, A.E., Berndt,
S.I., Lettre, G., Frayling, T.M., Okada, Y., Wood, A.R., Visscher, P.M.,
Hirschhorn, J.N., 2022. A saturated map of common genetic variants
associated with human height. Nature 610, 704--712.
https://doi.org/10.1038/s41586-022-05275-y

Zhao, Y., Hou, Y., Xu, Y., Luan, Y., Zhou, H., Qi, X., Hu, M., Wang, D.,
Wang, Z., Fu, Y., 2021. A compendium and comparative epigenomics
analysis of cis-regulatory elements in the pig genome. Nat Commun 12,
2217.

\end{document}